# Computational Modeling of Deep Multiresolution-Fractal Texture and Its Application to Abnormal Brain Tissue Segmentation

A. Temtam, L. Pei, and K. Iftekharuddin

*Abstract*—Computational modeling of Multiresolution-Fractional Brownian motion (fBm) has been effective in stochastic multiscale fractal texture feature extraction and machine learning of abnormal brain tissue segmentation. Further, deep multiresolution methods have been used for pixel-wise brain tissue segmentation. Robust tissue segmentation and volumetric measurement may provide more objective quantification of disease burden and offer improved tracking of treatment response for the disease. However, we posit that computational modeling of deep multiresolution fractal texture features may offer elegant feature learning. Consequently, this work proposes novel modeling of Multiresolution Fractal Deep Neural Network (MFDNN) and its computational implementation that mathematically combines a multiresolution fBm model and deep multiresolution analysis. The proposed full 3D MFDNN model offers the desirable properties of estimating multiresolution stochastic texture features by analyzing large amount of raw MRI image data for brain tumor segmentation. We apply the proposed MFDNN to estimate stochastic deep multiresolution fractal texture features for tumor tissues in brain MRI images. The MFDNN model is evaluated using 1251 patient cases for brain tumor segmentation using the most recent BRATS 2021 Challenges dataset. The evaluation of the proposed model using Dice overlap score, Husdorff distance and associated uncertainty estimation offers either better or comparable performances in abnormal brain tissue segmentation when compared to the state-of-the-art methods in the literature.

## I. INTRODUCTION

COMPUTATIONAL modeling is a powerful tool for predicting complex events and phenomena in the nature. It involves mathematical models, and corresponding computational methods to represent underlying physical processes. Computational modeling can play an effective role to segment and track tumor volume. Accurate segmentation and volume measurement of abnormal brain tissues, such as enhancing and non-enhancing tumors, edema, and necrosis, is crucial for early diagnosis, treatment planning, and patient survival prediction.

Heterogeneity among different types of abnormal tissues presents a challenge in robust brain tumor segmentation. Various methods including atlas-based [1, 2], feature-based machine learning [3-5] and deep learning [6-8] approaches have been proposed for brain tumor segmentation. The use of multiresolution fractional Brownian motion (fBm) has shown promise in estimating the abnormal brain tissue segmentation [9, 10].

Multiresolution techniques are important component of computational modeling that allow for flexible and efficient analysis of systems and signals at different scales or resolution. Computational models, such as fractal and multifractal analyses, have proven successful in predicting tumor growth and segmentation of brain tumor volumes among others [11, 12]. Fractal analysis in images represents texture [13], while multifractal models depict the scale-dependent propagation of voxel heterogeneity [14]. Various methods, including the estimation of the Hurst exponent, have been utilized to calculate the fractal dimension (FD) of texture [15] [16].

Deep learning (DL) methods such as Convolutional Neural Networks (CNN), on the other hand, have shown success in a wide range of image classification tasks [8, 17] . Mathematical analysis of features generated by deep convolutional neural networks has been discussed [8, 17] . Different variations of CNN models using basic U-Net [18] have shown success for medical image segmentation. The major drawbacks for basic U-Net, including the loss of spatial information [19-21] and difficulty in handling images with variations in lesion or tumor size [21], motion artifacts, low spatial, and temporal resolution [21], impose challenges to properly segment the region of interests. Though DL methods seek to overcome the lack of generalizability in feature extraction by using transfer learning, the need for large amount of disease-specific labeled patient data and the process to tune large number of parameters in DL may still make it difficult to obtain robust abnormal tissue such as brain tumor segmentation.

Wavelet convolutional neural network (wavelet-CNN), a deep multiresolution analysis method, has been shown to offer improved texture classification [22]. The authors in [13]

We acknowledge partial funding of this work by a NIBIB/NIH grant # R01 EB020683.

A. Temtam is with the Vision Lab, Electrical and Computer Engineering, Old Dominion University, Norfolk, VA 23529 USA (e-mail: atemt001@odu.edu).

L Pei., was with Vision Lab, Electrical and Computer Engineering - Old Dominion University, Norfolk, VA 23529 USA. He is now with the Frederick National Laboratory for Cancer Research (e-mail: lxpei001@odu.edu).

KM Iftekharuddin is with the Electrical and Computer Engineering - Old Dominion University, Norfolk, VA 23529 USA, (e-mail: kiftekha@odu.edu).

presents a multi-level wavelet-CNN for image restoration. The wavelet-CNN have also been used in medical image analysis for brain MRI Image classification for cancer detection [23]. Specifically, the authors analyze scattering networks, where signals are propagated through layers that compute semi-discrete wavelet transforms followed by modulus non-linearities. It is shown that the resulting wavelet-based feature extractor is stable [8, 17].

To the best our knowledge, there are no computational methods that mathematically models deep learning of multiresolution fractal texture features in the literature that may offer robust feature engineering. Consequently, this work proposes a novel modeling of Multiresolution Fractal Deep Neural Network (MFDNN) and its computational implementation that obtains mathematical model of multiresolution fBm and deep multiresolution analysis. The resulting MFDNN model offers the desirable property of automatically extracting stochastic multiresolution texture features by analyzing large amount of image data in a deep neural network framework. We further show an application of the proposed MFDNN for brain tumor segmentation using the most recent BRATS 2021 Challenge patient data [24-26]. The evaluation using the Dice overlap score, Husdorff distance and associated uncertainty estimation offers improved robust performance in abnormal brain tumor tissue segmentation when compared to the state-of-the-art methods in the literature.

The rest of the paper is organized as follows. Section II discusses the related work. Section III introduces the proposed MFDNN model. Section IV and V discuss the experiments used to evaluate the model performance and shows the corresponding results, Finally, Section VI presents discussion, conclusions, and future directions.

## II. MATHEMATICAL PRELIMINARIES

This section discusses the essential background for this work to include Fractal Dimension (FD), multiresolution estimation of Hurst exponent (H) and Multiresolution CNN (MCNN). The MCNN allows for multiresolution estimation of Hurst exponent and hence FD in a deep learning framework.

### A. Fractional Dimension (FD)

Fractal features represent texture parameters in texture images [27]. Fractal analysis is obtained using estimation of the FD which represents a universal description of the inhomogeneities in the image. Consequently, fractal geometry may be used for the characterization and segmentation of tissues in many medical imaging applications [28]. Theoretically, FD and Hurst exponent (H) are independent. FD of a shape or a surface is a measure of roughness with FD $\in$ [n, n + 1). FD is a measure for a surface in n-dimensional space with higher values representing rougher surfaces. For self-similar processes, FD and Hurst coefficient (H) are given as [29, 30],

$$FD = n + 1 - H. \qquad (1)$$

The models allow for the straightforward synthesis of images with arbitrary fractal properties and power-law correlations. A stationary, self-similar stochastic process such as fBm (e.g., the Gaussian process with correlation function) is defined as [31, 32],

$$B_H(t) - B_H(s) = \frac{1}{\Gamma(H+0.5)} [\{ \int_{-\infty}^{0} (t-s)^{H-0.5} - s\,H - 0.5\,dB(s) + \int_{0}^{t} (t-s)^{H-0.5}\,dB(s) \}]. \qquad (2)$$

Here $s$ is an ordinary Brownian motion, H is the Hurst Coefficient, and fBm is non-stationary process.

To obtain a combination of, multiresolution H with wavelet as introduced in [6], and the wavelet transform of x(s) is defined as,

$$W_x(s, a) = \langle x(T), \psi_{s,a} \rangle = \int x(T)\,\psi_{s,a}(T)dT. \qquad (3)$$

Note that the continuous wavelet transform [33, 34] is defined as,

$$\psi_{s,a}(T) = \frac{1}{|a|^{0.5}}\,\psi_{s,a}\left(\frac{(T-s)}{|a|}\right), \qquad (4)$$

where $\psi_{s,a}(T)$ is obtained by the affine transformation of the mother wavelet with $s, a \in R$, $a \neq 0$. Using time signal wavelet scale convolution, we can write (4) as,

$$W_x(s, a) = x(s) * \psi_{s,a}(T). \qquad (5)$$

The expected value of the squared-magnitude of the wavelet transform as in [35] is given as,

$$E[|W_x(s, a)|^2] = \int |x(s) * \psi_{s,a}(T)|^2 dt_{s,a}. \qquad (6)$$

Finally, if $X$ is a self-similar process with stationary increments, the scattering moments of fBm can be written as [36],

$$\{|X * \psi_{j1}(t)|\} \approx \{2^{j_1 H}\}. \qquad (7)$$

Taking the log on both sides yields,

$$\log\{|X * \psi_{j1}(t)|\} \approx Hj_1 \log\{2\}. \qquad (8)$$

And,

$$H_{j1} \approx \frac{\log\{\langle |X * \psi_{j1}(t)|\rangle\}}{(\log 2)}. \qquad (9)$$

The estimate of FD than may be obtained using Equation (9) and (1).

### B. Multiresolution Convolutional Neural Networks (MCNN)

Convolutional Neural Networks (CNNs) have robust adaptive learning capabilities. Several recent works [37] demonstrate that a CNN in combination with different multiresolution framework may offer much better accuracy for segmentation than using a CNN only. multiresolution is a good multiscale feature extraction tool because of the solid and formal mathematical framework [38]. MCNN combines the multi-resolution wavelet transform and convolutional neural networks to improve the performance of the networks [39]. For

 

a given function , a MCNN simply obtains the multi-resolution wavelet decomposition framework as discussed in [40], and the correspondence between the input and output of the convolutional neural network is defined as [41],

$$y_i = f_i \sum_{i=1}^{n} \overline{\psi}_i \, x_i + bi, \tag{10}$$

where $x$ is the input, $y_i$ is the output, $\overline{\psi}_i$ is the convolution matrix of the $i$-th layer includes the bias by having a constant input, and $bi$ is the deviation of the $i$-th layer.

Each layer identifies the weights $\overline{\psi}_i$ as coefficients over $i$-th. By combine parameters, CNNs reduce the total of parameters and accomplish translation invariance in the image space. The definition of $y_i$ in Equation (10) corresponds to convolution of $x_i$ through a filtering kernel $\overline{\psi}_i$, therefore this layer is called a convolution layer. $y_i$ in Equation (10) can be written as convolution expression,

$$y = X * \overline{\psi} \tag{11}$$

Equation (11) shows convolution and pooling in CNNs as filtering and down sampling operations. This formulation allows to connect convolution and pooling with multiresolution analysis. Hence, Equation (11) can be written using a Hankel matrix that is related to the convolution operations in CNN as follows [42, 43],

$$y = x * \overline{\psi} = H_d(x)\psi \in R^n. \tag{12}$$

For a more detailed CNN convolution operation in the form of Hankel matrix, and a multi-resolution analysis for convolution framelets using wavelet see [42, 43].

### C. Multiresolution Analysis via Deep Convolutional.

In deep convolutional networks, the local convolution filters play a crucial role in achieving effective reduction behavior. However, non-local basis is also an important design parameter that greatly affects the performance of the network. In this work we incorporate non-local basis with local convolution filters in the decoder that utilizes wavelet transformation. The wavelet transform decomposes the input image into multiple frequency subbands, which are then used to reconstruct the image more efficiently.

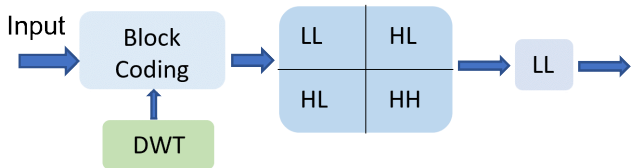

Fig .1. Structure of the Multiresolution Encode

To obtain a combined feature representation, the multiresolution coefficients are concatenated with the computed fractal dimension. The resulting feature representation is then passed through a series of convolutional layers, similar to a basic U-Net. This step restores the spatial resolution of the feature maps and increases the number of channels. In the decoder, an inverse wavelet transform is applied to the final feature maps to obtain the segmentation mask. This approach captures both local and non-local information effectively, leading to improved performance in computational modeling tasks, such as brain tumor segmentation.

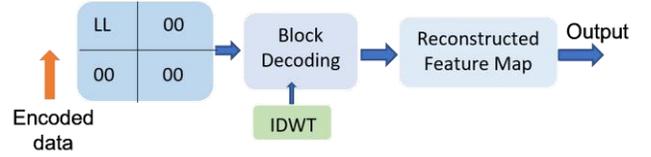

Fig .2. Structure of the Multiresolution Decode

### III. METHOD

This section discusses the proposed mathematical model and the corresponding computational implementation.

### A. Mathematical Model

Let us consider the multiresolution convolution operation non-linearity $(M_n)$ followed by pooling operation $(P_n)$ to combine MCNN to FD. In order to obtain the empirical estimate of the $q$-th order moment of $|W_x(s, a)|$ [35] [44], the expected value is defined as:

$$E\{|W_x(si, a)|^q\} = \left(\frac{1}{N}\right) \sum_{i=0}^{N-1} |W_x(si, a)|^q. \tag{13}$$

This estimation is based on that assumption that the wavelet series $Wx(s, a)$ comply with the normality and stationary properties within scale. Then we replace the wavelet-modulus operation $W_x(s, a)$ by a convolution with the wavelets and squared the expected value as follows,

$$E[|X|^2] = E\left[|X * \psi_j(t)|^2\right]. \tag{14}$$

Next, we replace the $q$-th order by 2 in (13), and also replace the operation $W_x(s, a)$ by $X$ as shown in (14) to introduce a non-linearity $(M_n)$[45]. Note $M_n$ satisfies the Lipschitz property given as,

$$\|M_n X - M_n \psi_{s,a}(T)\|_2^2 \le \text{Ln} \| X - \psi_{s,a}(T) \|_2^2. \tag{15}$$

where, $M_n X = 0$ for $X = 0$. The output of this non-linearity, in (15), is then pooled using a pooling function, $(P_n)$, [45] yielding the operator associated with n-th layer network as follows,

$$f \to S_n d/2 \, P_n \, (M_n \, (X * \psi_j(t)) \, (S_n)); \tag{16}$$

and

$$f \to S_n(d/2) P_n \, (M_n\{\left(\frac{1}{N}\right) \sum_{i=0}^{N-1} |W_x(si, a)|^2 \}; \tag{17}$$

where, $P_n$ is Pooling and Sn $\ge 1$ is the pooling factor. Note $P_n$ satisfies the Lipschitz property given as,

$$\|P_n X - P_n \psi_j(t)\|_2^2 \le \text{Rn} \| X - \psi_j(t) \|_2^2 \tag{18}$$
$$\text{for all } X, \psi_j(t);$$

---

**Algorithm 1:** MFDNN Training Model.

---

1: /* Initialization */

---

**Input:** Define $X$(mMRI image) $X \in R^{N \times H \times W \times C}$ . & corresponding $GT$

    -Load the Volume dataset

---

2: /* Initialize the Network/*

---

-Define the U-Net architecture with encoder and decoder.
-Define the loss function.
-Compile the model with an optimizer.
- Set Wavelets, $\psi_j(t)$.

---

3-/*Train the Network/*

---

**for** each $X$**do:**

    Initialize the coefficient $f_i$;

    Apply $\psi_i(t)$ on Input $X$;

      Compute the $FD$ of $X$.

    **for** each $i$ **do:**

      obtain feature maps $Z_i = f(X_{fractal-wavelet} * F_i + b_i)$

      Restore the spatial resolution $X_{upsampled_i} = upsample(Z_{bottlenec}, k)$

      Obtain $Z_i = f(X_{upsampled_i} * F_i + b_i)$

      Obtain $\hat{Y} = softmax(Z_i * F_{out} + b_{out})$

    **end**

**end**

**return**

---

Fig 3 MFDNN Algorithm

where $P_n X = 0$ , for $X = 0$, for all X, $\psi_j(t)$.

Plugging in (17) into (18), yields the pooling factor associated with output of n-th network layer is given as:

Now, let $y = S_n x$, then we have,

$$f \rightarrow \int |S_n d/2 \, P_n \, (M_n \, (X * \psi_j(t)) \, (Snx)|^2 \, dx_{s,a}$$
$$= S_n d/2 \int | P_n \, (M_n \, (X * \psi_j(t)) \, (S_n x)|^2 \, dx_{s,a} . \quad (19)$$

Further, let $y = S_n x$, then we have,

$$dy = S_n \, dx = \int | P_n \, (M_n \, (X * \psi_j(t)) \, (y)|^2 \, dy_{s,a}. \quad (20)$$

As shown in [45] the specific form of the Lipschitz non-linearities and Lipschitz pooling operators satisfy (20), and the scattering moments of fBm [36], respectively. If $X$ is a self-similar process with stationary increments [45], we get:

$$\{|X * \psi_j(t)|\} = P_n \, (M_n\{2^{j \, H}\{X * \psi_j(2^j t)\}, \quad (21)$$

and,

$$P_n \, (M_n\{2^{j \, H}\{X * \psi_j(2^j t)\} \approx P_n \, (M_n\{2^{j \, H} \}). \quad (22)$$

Taking the log on both sides of (22) yields,

$$H_j = \lim_{j \rightarrow 0+} \frac{\log\{P_n \, (M_n \, (|X * \psi_j(t)|\}}{P_n \, (M_n \, j(log_2 2)}; \quad (23)$$

where j= 1,2,.. , $* \, convloution, M_n$ non-linearity and $P_n \, is \, Pooling$. For self-similar time series, H is directly related to fractal dimension, FD, as defined in (1) where The parameter n is Euclidean dimension 1 for 1-D fractional Brownian motion (2 for 2-D, 3 for 3-D and so on) of the space.

### B. Algorithmic Modeling

The proposed MFDNN computational modeling algorithm pseudo code is shown in Fig 3. The proposed framework utilizes a basic U-Net [46] [47] architecture combined with multiresolution, and fractal dimension Fig 4. The architecture contains multiresolution encoding module, a fractal dimension (FD) context encoding module, and multiresolution decoding module. The network consists of a contracting path and an expansive path. The contracting path reduces the data while increasing the feature information through repeated convolution and pooling operations, followed by a rectified linear unit (ReLU) activation. The expansive pathway combines the feature and spatial information through up-convolutions and concatenations. The input image for the

 

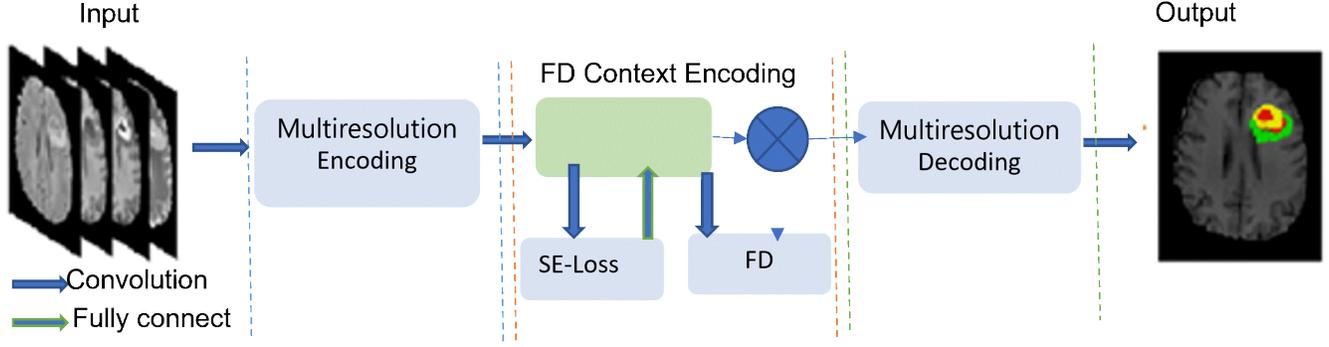

Fig 4. Proposed computational model for MFDNN

MFDNN is decomposed into multiple frequency subbands as follows,

$$X_{wavelet} = wavelet_{transform}(X). \quad (24)$$

The fractal dimension of the image is computed and concatenated with the wavelet coefficients to form a combined feature representation. The fractal dimension computed using Equation (1). The resulting combined feature representation is represented as,

$$X_{fractal-wavelet} = \quad (25)$$
$$concatenate(fractal_{dimension(X)}, X_{wavelet}).$$

The combined feature representation is passed through a series of convolutional layers, similar to a basic U-Net. The feature maps $Z_i$ are obtained as:

$$Z_i = f(X_{fractal-wavelet} * F_i + b_i); \quad (26)$$

where f is the activation function, $b_i$ is the bias term, and * represents the convolution operation. Max-pooling layers are used to reduce the spatial dimensions of the feature maps:

$$X_{pooled_i} = max\_pool(Z_i, k). \quad (27)$$

The bottleneck layer reduces the number of channels and retains the features learned from the previous layers. It is represented as:

$$Z_{bottleneck} = f(X_{pooled_i} * F_{bottleneck} + \quad (28)$$
$$b_{bottleneck}).$$

The decoder path restores the spatial resolution of the feature maps and increases the number of channels. The upsampled feature maps are obtained as:

$$X_{upsampled_i} = upsample(Z_{bottlenec}, k). \quad (29)$$

The convolutional layers are applied to the upsampled feature maps, resulting in feature maps $Z_i$ :

$$Z_i = f(X_{upsampled_i} * F_i + b_i) \quad (30)$$

The final feature maps are passed through an inverse wavelet transform to obtain the final segmentation mask:

$$\hat{Y} = inverse\_wavelet\_transform(Z_i) \quad (31)$$

The final layer of MFDNN is a 1x1 convolutional layer that maps the feature maps to the desired number of classes, typically two in the case of tumor segmentation:

$$\hat{Y} = softmax(Z_i * F_{out} + b_{out}) \quad (32)$$

The loss is still computed between the predicted and ground truth masks, represented by $\hat{Y}$ and Y, respectively:

$$L = -(1/N) * sum(Y * log(\hat{Y}) + \quad (33)$$
$$(1 - Y) * log(1 - \hat{Y})).$$

The network is trained using the Adam optimizer that is used to minimize the loss and update the network parameters.

## IV. DATASETS

### A. Brain Tumor Segmentation Dataset

We use Multimodal Brain Tumor Segmentation BraTS 2021 Challenge dataset with 1251 patient multi-modality MRI (mMRI) cases [26]. For each patient case in this dataset, all mMRI sequences (T1, T1ce, T2, and T2-FLAIR) are considered. These mMRIs are preprocessed by the Challenge organizers, and the pre-processing steps include co-registration, noise reduction, and skull stripping. The preprocessed image size is $240 \times 240 \times 155$. Ground truth is labeled as 1, 2, 4, 0 for NC, ED, ET, and everything else, respectively. The evaluation performances are based on three tumor subregions: whole tumor (WT), tumor core (TC), and enhancing tumor (ET). The WT is the combination of all tumor tissue types including ED, TC and ET. The TC consists of NC and ET, while ET is the enhancing tumor only. We first build a MFDNN model using the BraTS 2021 training data and then validate the model using the validation dataset. Finally, we apply our validated model to unknown test data, obtain the WT, TC and ET segmentation results and submit these results to the official Challenge portal for online evaluation.

## V. EXPERIMENTS

### A. Implementation Details

The proposed MFDNN is trained end-to-end on a dataset of brain scans in MRI. The input to the network is mMRI represented by a matrix X of size H x W x C ($240 \times 240 \times$

155) with 4 channels and Ground truth. The MRI images is first passed through multiresolution. The framework is implemented under the open-source deep learning library PyTouch. In the training phase, the learning rate is initially set as 0.001 and decreased by a weight decay. The momentum is 0.9, and due to the limitation of the memory, we choose 1 as the batch size. Experiments are carried out on GTX1080 GPU with 8 GB of video memory and the CUDA edition is 10.0. The fractal dimension (FD) context encoding module, captures global texture features using a semantic encoding loss (SE-loss).

We perform data normalization of all cases to have mean square error (NMSE) value, which is defined unit standard deviation as in the Equation (34).

$$NMSE = \frac{\sum_{i=1}^{M}\sum_{j=1}^{N}[f^*(i,j) - \hat{f}(i,j)]^2}{\sum_{i=1}^{M}\sum_{j=1}^{N}[f^*(i,j)]^2}, \quad (34)$$

where $\hat{f}$ and $f^*$ denote the reconstructed images and ground truth, respectively. M and N are the number of pixels for row and column. As the original dimension of each modality is 240 x 240 x155, to manage computational memory and cost. The size of the images is reduced by cropping to a dimension of 192x160 x128.

The most important aspects of the proposed MFDNN are the FD context encoding module, which computes FD and the scaling factors related to the representation of all classes. These factors are learned simultaneously in the training stage. These factors capture the global information of all classes essentially learning to alleviate the training preference that may occur due to imbalanced class representation in data. To calculate the Semantic Error (SE-loss), we construct another fully connected layer with sigmoid activation function upon the encoding layer such that predict object segmentation in the image [12]. The encoding module extracts high-dimensional features from the input, while the FD context encoding module produces updated features and a semantic loss to ensure all segmentation classes are represented.

### B. Performance Evaluation

The performance of the models is evaluated using online submission BraTS 2021 challenge evaluation portal. The online evaluation uses two metrics [26]: Dice score and the 95th percentile of the symmetric Hausdorff distance (HD95). Both metrics are evaluated over the whole tumor (WT), core tumor (TC), and enhancing tumor (ET) sub regions. The Dice score Equation (35) is used to evaluate the quality of automated segmentations and to demonstrate both size and localization agreement compared to the ground truth given as,

$$\text{Dice} = \frac{2TP}{FP + 2TP + FN}. \quad (35)$$

On the other hand, the HD95 for accuracy of segmentation computation is given as,

$$HD95 = 95\% \ (d \ (Y, \hat{Y}) \| d(Y, \hat{Y})) \quad (36)$$

where d is the element-wise distance of every voxel Y to the closest voxel of the same label in the second, $\hat{Y}$ are the segmented of each voxel, Y is the GT of each voxel, and ‖ is the concatenation operator. The HD95 measures the distance between the predicted and ground truth tumor boundaries and provides a measure of how well the segmentation algorithm can capture the precise shape and location of the tumor. This metric reflects the clinical significance of segmentation errors. The brain tumor tissue sgmentation performance for MFDNN is shown in the Table I. Overall, our approach reached an improved result in the whole tumor (WT) segmentation task with mean Dice score of 91.33 % for the online BraTS 2021 validation dataset. In most volumes the Dice coefficients of the model shows segmentation of the Tumor Core (TC) with mean Dice score of 84.95 % for the validation dataset. Fig .7. shows the Dice score for Whole Tumor (WT), Enhanced Tumor (ET), and Tumor Core (TC) contours for each patient using the validation dataset.

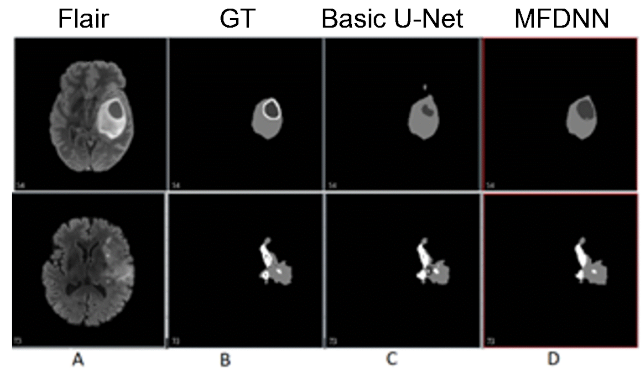

Fig 5. Abnormal brain tissue segmentation results: (A) Flair MRI image, (B) Ground truth, (C) Basic U-Net, (D) MFDNN; Each row represents an Input MRI modality:(A) Flair, (B) T1, (C) T1C, (D) T2

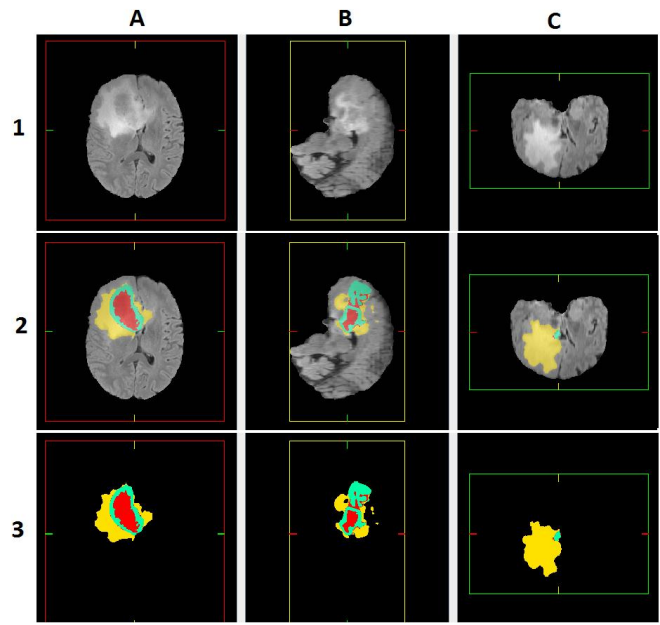

Fig 6. MFDNN-MCDO segmentation results of brain tumor, (A) Flair Image (B) Sagittal View (C) Coronal View from Testing set, which is part of BraTS /2021 Validation dataset Challenge. tumor core is visualized in yellow, enhancing tumor is visualized in red, and is tumor core visualized in Green, (3) shows the Segmentation results.

TABLE II Online evaluation performance comparison Performance for three models on the BraTS 2021 VALIDATION dataset for brain tumor segmentation.

| MFDNN Model | Statistical parameter | Dice Score | | | HD95 | | |
|---|---|---|---|---|---|---|---|
| | | ET | TC | WT | ET | TC | WT |
| MFDNN | mean | 0.798093 | 0.849537 | 0.913389 | 21.23622 | 8.180387 | 4.264839 |
| | sd | 0.232866 | 0.213613 | 0.076908 | 80.97778 | 36.00219 | 5.899107 |
| | median | 0.864205 | 0.92991 | 0.937525 | 2 | 2 | 2.44949 |
| | 25quantile | 0.796387 | 0.849334 | 0.891405 | 1.414214 | 1.103553 | 1.732051 |
| | 75quantile | 0.913061 | 0.958211 | 0.956281 | 3 | 4.898979 | 4.527356 |
| MFDNN-MCDO | mean | 0.802106 | 0.848733 | 0.913673 | 16.37481 | 8.204463 | 4.23459 |
| | sd | 0.224283 | 0.215755 | 0.07639 | 69.56121 | 36.00857 | 5.890024 |
| | median | 0.864701 | 0.929893 | 0.937363 | 2 | 2 | 2.44949 |
| | 25quantile | 0.796735 | 0.848963 | 0.891342 | 1.414214 | 1 | 1.493673 |
| | 75quantile | 0.911981 | 0.957783 | 0.956326 | 2.914214 | 4.94949 | 4.582576 |
| MFDNN Ensample | mean | 0.801473 | 0.846156 | 0.910879 | 13.2448 | 8.20279 | 4.565624 |
| | sd | 0.214224 | 0.214661 | 0.076828 | 60.52547 | 35.9914 | 7.109701 |
| | median | 0.86549 | 0.929881 | 0.935277 | 2 | 2 | 2.828427 |
| | 25quantile | 0.794222 | 0.848023 | 0.890366 | 1.414214 | 1.414214 | 1.732051 |
| | 75quantile | 0.91121 | 0.954052 | 0.95342 | 3 | 5.04951 | 4.582576 |

The MFDNN outperforms the basic U-Net in terms of the Dice coefficient for all three subregions of the brain tumor. The proposed method also achieves lower HD95 distances for ET and TC when compared to UNET, indicating that the proposed method produces more precise segmentations with fewer boundary errors in these subregions of brain tumor. We implement MFDNN-MCDO model that achieves lower Hausdorff95 distances for ET and TC compared to the MFDNN method, indicating that the proposed method produces more accurate segmentation results with fewer boundary errors in these subregions. The MFDNN-MCDO achieves 4.28% lower

Dice coefficient for ET, 2.28% lower for TC and 1.43% lower for WT than that of MFDNN, respectively.

Overall, our method performs better than the best top methods in terms of the HD95 distance for ET and TC, but performed slightly less in terms of the HD95 distance for WT. Moreover, the proposed method exhibited superior performance compared to the basic U-Net model for both the Dice coefficient and HD95 distance for ET and TC, while achieving similar results for WT.

### C. Uncertainty Analysis

Uncertainty measures in deep learning are essential for understanding the reliability of a model's predictions and assigning a level of confidence to model predictions. There are several methods available for estimating the uncertainty of a deep learning model. We obtain two different methods for uncertainty quantification as discussed in the next section.

### i. MFDNN with Monte Carlo Dropout

Monte Carlo Dropout (MCDO) is a technique that uses dropout layers in DNN architectures to create variation in the output of the model outputs to compute the prediction uncertainty.

We adopt this technique for the MFDNN by activating dropout layers in MFDNN. We then repeat each testing samples y, N times at during testing and pass each image through the dropout that we turn on during the evaluation. Next obtain the N different predictions $P_N(y)$. that we turn on during the evaluation. Next obtain the N different predictions $P_N(y)$ as follows,

$$P_N(y) = \frac{1}{N} \sum_{n=1}^{N} P_n(y). \qquad (37)$$

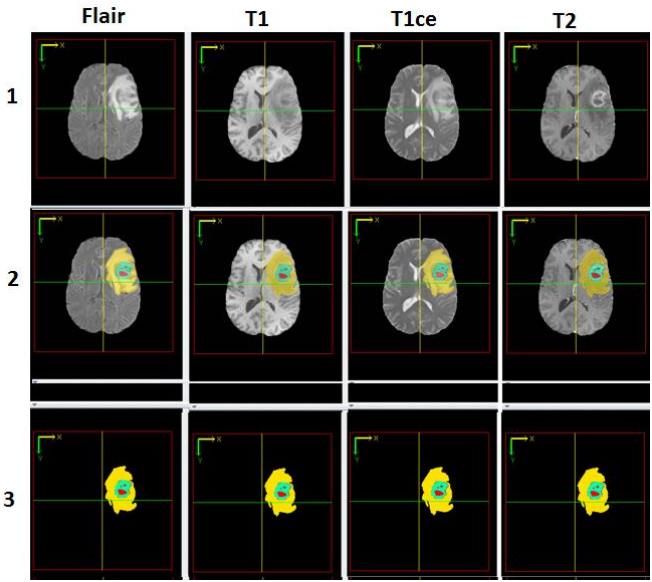

Fig. 7. MFDNN-Ensample Segmentation results of brain tumor on. (1) BraTS 2021 Validation Data, (2,3) Segmentation results of Brain tumor.

TABLE II Online evaluation performance comparison with different methods on the same BraTS 2021 VALIDATION dataset for brain tumor segmentation
.

| | Dice | | | HD95 | | |
|---|---|---|---|---|---|---|
| Method | ET | TC | WT | _ET | _TC | _WT |
| S. Roy, et al. 45 | 84.09 | 87.85 | 92.77 | 22.41 | 9.2 | 3.42 |
| A. S. Akbar, et al.46 | 77.73 | 82.19 | 89.33 | 30.90 | 16.09 | 11.34 |
| J. Roth, et al. 47 | 0.79185 | 0.83494 | 0.90638 | 16.60631 | 10.11467 | 4.53913 |
| C. Saueressig, et al. 48 | 0.734 | 0.807 | 0.894 | 28.20 | 12.62 | 6.79 |
| Y. Choi, et al. 49 | 0.7321 | 0.7514 | 0.8743 | 35.0074 | 24.6376 | 10.1613 |
| M. M. Rahman, et al. 50 | 0.7696 | 0.8281 | 0.9121 | 20.24 | 12.10 | 4.71 |
| Basic U-Net 16 | 0.762893 | 0.794146 | 0.91269 | 27.49043 | 16.25987 | 4.604289 |
| MFDNN | 0.798093 | 0.849537 | 0.913389 | 21.2362 | 8.18038 | 4.26483 |
| MFDNN-MCDO | 0.802106 | 0.848733 | 0.913673 | 16.37481 | 8.204463 | 4.23459 |
| MFDNN-Ensample | 0.801473 | 0.846156 | 0.910879 | 13.2448 | 8.20279 | 4.565624 |

In order to obtain epistemic uncertainty in the predictions, we perform test time augmentation (TTA) of inputs, to calculate the average prediction score.

*ii.* *MFDNN with Deep Ensembles*

In Deep Ensembles, we randomize initial weights and obtain multiple MFDNN models trained on the same training data then construct multiple best performing MFDNN models, each with their own set of parameters, denoted as $\{\theta_1, \theta_2, ... \theta_n\}$. Each model in the ensemble makes a prediction, denoted as $\{y_1 (x; \theta_1), y_2 (x; \theta_2), ...., y_n (x; \theta_n), \}$[48], where x is the image input data and $\theta_i$ is the set of parameters for the i-th model. The final prediction is obtained by averaging the predictions of all the models in the ensemble. This can be represented as:

$$y(x) = \frac{1}{N} \sum_{i=1}^{N} y_i(x, \theta_i). \tag{38}$$

*D. Results and discussion*

In this section, the MFDNN model, Monte Carlo, and deep ensemble methods are used to estimate Dice coefficient and HD95. The Monte Carlo method provides the highest level of accuracy in EN and TC compared to the MFDNN model and deep ensemble, however, for WT (0.913) all three methods are comparable. For uncertainty estimation, the Monte Carlo method, can provide a more uncertain prediction by sampling from the model or its parameters as it captures the model's stochastic nature. The deep ensemble method can provide a more robust and uncertain prediction than a single model by averaging the predictions of many models. Therefore, the Monte Carlo method may be more suitable for applications that requires uncertain predictions, whereas the deep ensemble method may be more suitable for applications that require robust predictions.

Further, the Monte Carlo method is less computationally expensive than the deep ensemble method. The MFDNN model is the least computationally expensive, however, it lacks robustness and uncertainty measure. Table II shows that MFDNN, MFDNN-MCDO, and MFDNN-Ensample offer either better or comparable performance when compared to several other recent works for the Dice scores across all tumor sub-regions (ET, TC, and WT). Further, for HD95 values, MFDNN-MCDO offers the lowest values, indicating that it has the most accurate segmentation performance. Therefore, MFDNN-Ensample is the best performing algorithm overall for the three versions of the proposed methods. [49] [50] [51] [52] [53] [54]

## VI. Conclusion

This study proposes a novel mathematical and computational model for MFDNN that offers the strengths of multiresolution-Fractional Brownian motion (fBm) modeling and deep multiresolution analysis. The MFDNN is used to estimate the FD and to characterize the multiscale texture properties of different types of abnormal tissues in MRI. The MFDNN model is further enhanced by implementing uncertainty-aware MFDNN-MCDO and MFDNN- Ensemble models enabling more accurate and objective quantification of the tumor disease burden. The efficacy for all three models is evaluated using over 1200 patient cases from the most recent BRATS 2021 Challenge dataset. The evaluation results show that all three proposed MFDNN models in this study outperform the state-of-the-art methods considered in this study using the Husdorff score, while only [45] performs better than the proposed methods using Dice score for brain tumor volume segmentation. This work highlights the potential of computational multiresolution texture modeling to improve the accuracy and reliability of automated abnormal brain tissue segmentation, which can contribute to early diagnosis and more effective treatment planning for diseases such as brain tumors. The performance of the MFDNN may be evaluated further using multiple patient datasets from different centers. In addition, further work could involve extending the proposed model to various other tasks related to abnormal tissue segmentation, beyond those examined in the current study.